\documentclass[journal,12pt,onecolumn]{IEEEtranTCOM}
\ifCLASSINFOpdf
\else
\fi

\usepackage{longtable}
\usepackage{amsmath}
\usepackage{cancel}
\usepackage{amssymb}
\usepackage[dvips]{graphicx}
\usepackage{multirow}
\usepackage{array}
\usepackage[justification=centering]{caption}
\usepackage[labelsep=period]{caption}
\usepackage[font=footnotesize]{caption}
\usepackage{setspace}
\usepackage{float}
\usepackage{mathrsfs}
\usepackage{algorithm}
\usepackage{algpseudocode}
\usepackage{algcompatible}
\usepackage{color}

\usepackage{etoolbox}
\patchcmd{\subequations}{\def\theequation{\theparentequation\alph{equation}}}%
{\def\theequation{\theparentequation.\arabic{equation}}}{}{}

\begin{document}
	
\title{On the Feasibility of Interference Alignment in 	Compounded MIMO Broadcast Channels with Antenna Correlation and Mixed User Classes}
	
\author{Galymzhan Nauryzbayev,~\IEEEmembership{Member,~IEEE,}~Emad~Alsusa,~\IEEEmembership{Senior~Member,~IEEE,} \\
	and~Mohamed Abdallah,~\IEEEmembership{Senior~Member,~IEEE}}
	
	
	\maketitle
	
\thispagestyle{empty}
	
\begin{abstract}
	This paper presents a generalized closed-form beamforming technique that can achieve the maximum degrees of freedom in compounded multiple-input multiple-output (MIMO) broadcast channels with mixed classes of multiple-antenna users. The contribution is firstly described within the context of a three-cell network and later extended to the general multi-cell scenario where we also show how to determine the conditions required to align the interference in a subspace that is orthogonal to the one reserved for the desired signals. This is then followed by an analysis of the impact of antenna correlation for different channel state information acquisition models. The proposed scheme is examined under both conventional and Large-scale MIMO systems. It will be shown that the proposed technique enables networks with any combination of user classes to achieve superior performance even under significant antenna correlation, particularly in the case of the Large-scale MIMO systems.
\end{abstract}
	
\IEEEpeerreviewmaketitle
	
\section{Introduction}

\IEEEPARstart{I}{nterference} alignment (IA) is a potential capacity-achieving technique in interference-limited networks, that was initially proposed in \cite{Cad_Jaf_IC} to establish additional degrees of freedom (DoF) in a single-input single-output (SISO) model, where it was shown that the optimal DoF is $\frac{K}{2}$ in $K$-user time varying interference channel (IC). The concept of IA (is concerned with) aligning all interfering signals into a specific subspace at the receiver side in order to exclusively reserve (a linearly independent) subspace for interference-free data transmission. In a relatively short time, the IA concept has attracted a significant amount of the researcher attention and a number of algorithms have been proposed for multiple-input multiple-output (MIMO) systems \cite{feasibilityMIMO_IA}-\cite{galym7}, \cite{CJS}-\cite{Paula}, to name a few.
	
The feasibility conditions of IA for MIMO IC were analysed in \cite{feasibilityMIMO_IA} and \cite{Bresler}. In \cite{Gou}, the authors considered the case of $K$-user $M\times N$ MIMO network. It was shown that the achievable total number of DoF equals $K\textrm{min}(M,N)$, if $K\le R$, and $\frac{RK}{R+1}\textrm{min}(M,N)$, if $K>R$; $M$ and $N$ are the number of transmit and receive antennas, respectively, and $R = \frac{\textrm{max}(M,N)}{\textrm{min}(M,N)}$. The authors in \cite{mmk} investigated the X-channel network scenario and obtained remarkably high DoF compared to the findings in \cite{Jaf_F}. The derived outcomes of two DoF per user, \cite{mmk}, \cite{MMK}, prompted the work in \cite{Jaf_Sha}, where the authors presented an IA scheme for the two-user X-channel with three antennas at each node. In \cite{Cad_X}, the authors achieved a total DoF of $\frac{MN}{M+N-1}$ for the case of $M\times N$ time-varying X-channels. In \cite{tse}, the authors proposed an IA scheme to estimate the achievable DoF in cellular networks which is based on aligning the interference into a multi-dimensional subspace at multiple non-intended base stations (BS). On the other hand, in \cite{kim}, the authors provided a precise expression of the spatial multiplexing gain for two mutually interfering MIMO broadcast channels (BCs) using a linear transceiver. Moreover, for the same scenario, the authors in \cite{shin} proposed a novel IA technique for jointly designing the transmit and receive beamforming (BF) vectors with a closed-form non-iterative expression. It was shown both analytically and numerically that the proposed scheme achieves the optimal DoF. Furthermore, the authors in \cite{jie} showed how the extension of the grouping method, \cite{shin}, can be outperformed in terms of capacity and BS complexity. On the other hand, the authors in \cite{galym} considered a network scenario comprising all the aforementioned cases and proposed a closed-form IA scheme that can mitigate such mixture of interference. However, due to the limited available physical space, resulting in small separations between the antenna elements, each channel is affected by antenna correlation which hinders further enhancements of the transmission rate. 

\subsection{Main Contributions}
In contrast to \cite{galym} and \cite{galym1}, this paper considers the compounded MIMO BC network scenario when the users have different number of antennas and require different numbers of data streams. The derived algorithm can be regarded as a generalization that can be also used in the special cases of \cite{galym} and \cite{galym1}, where it was assumed that all users demand an equal number of data streams. Thus, this paper focuses on the feasibility of IA, achievable sum rate and DoF region under various scenarios of channel state information (CSI) acquisition and spatial correlation. The efficient channel model is used to capture all possible cases of CSI mismatch. Furthermore, the paper investigates the impact of spatial correlation between the antenna elements which arises due to the physical limitations of the available space. With this in mind, the model was manipulated to involve both the impact of spatial correlation and CSI mismatch in the compounded MIMO BC when the users have various antenna configurations and require to decode different numbers of data streams. Therefore, a comprehensive algorithm is proposed to define the minimum required antenna configuration to achieve a given DoF. The proposed technique is demonstrated within the contexts of both conventional and Large-scale MIMO (LS-MIMO) systems under a wide range of channel conditions and spatial correlations. Finally, the complexity of this technique is studied and compared to a well-known benchmark technique. 

The results demonstrate the effectiveness of the proposed technique particularly under highly correlated channels. It is shown that this technique is not only associated with relatively lower computational requirement, but also can achieve the maximum DoF even when the multiplexed users belong to different data classes.
	
\section{System Model}
\begin{figure}[h]
\centering
\includegraphics[width=0.5\columnwidth]{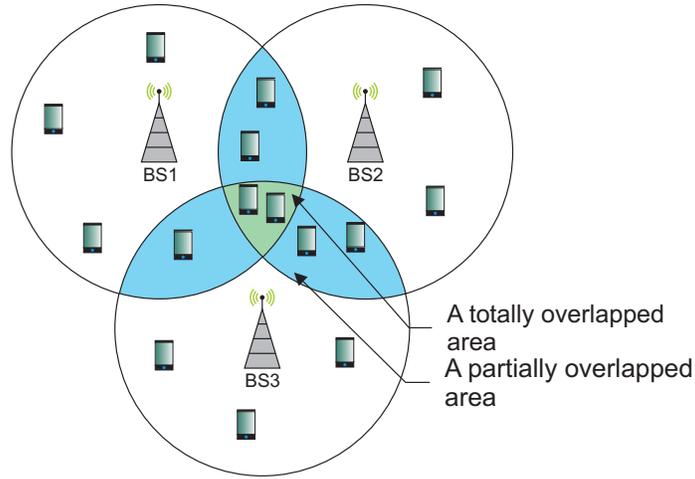}
\caption{A three-cell MIMO network with randomly distributed users.}
\label{fig:the_general_scheme}
\end{figure}

The system model considered here presents a cellular network comprising of $L$ cells serving multiple users randomly distributed as shown in Fig. \ref{fig:the_general_scheme}. In terms of the network coverage, the whole combined area can be determined as non-overlapped and overlapped areas, \cite{Alcaraz}, \cite{Zhao}. The overlapped area denotes a space where BSs cover a common region. Hence, the number of BSs creating the overlapped area varies from 2 to $L$. According to this number of BSs, we can further define totally and partially overlapped areas. Any user located in the totally overlapped area experiences the worst scenario when high interference degrades the network performance. Thus, a main focus of the work is the users located in the totally overlapped area. 

Under a dense network scenario, \cite{LS_survey}, it is feasible to assume that each BS aims to communicate with a large number of users $K$. We can also assume that $\mathcal{K}\le K$ users reside in the totally overlapped area, and that $\mathcal{K}$ is identical for all $L$ cells.

Since a concept of the point-to-point MIMO model can be applied for designing the cellular network, we consider the IC and X-channel models that are differentiable from each other by the message set ups, \cite{Cad_Jaf_IC}, \cite{CJS}, e.g., IC is the scenario when each BS $i$ serves only the corresponding user $i$ while the X-channel scenario is characterised by a message set when each BS $i$ has different messages to all users. Therefore, a cellular MIMO network can be determined as a compounded IC-X channel scenario when each cell consists of cell-edge users experiencing a multi-source transmission from $\mathcal{L}$ BSs such that $1 < \mathcal{L} < L$, and $\mathcal{L}$ is thought to be the same for all users. Hence, the user of interest can classify the observed network as a set of $(L-\mathcal{L})$ IC and $\mathcal{L}$ X-channels, and then the corresponding interferences can be defined as ICI and XCI, \cite{galym}, \cite{galym1}. 

Since we focus on multi-antenna users in the totally overlapped area, it is reasonable to assume that the user of interest may wish to receive data from several BSs. In turn, those BSs might send different numbers of data streams to the interested users. In this work, we consider the network scenario when BSs serve mixed user classes that can be determined as the case when users with different numbers of antennas require a different number of data streams to be decoded. Hence, we define the network configuration as $(M_{i},N_{i},\mathcal{K},L,d_{i})$, where $M_{i}$ and $N_{i}$ indicate the number of transmit and receive antennas per node in cell $i$, respectively, $\mathcal{K}$ is the number of users per cell in the totally overlapped area, and $d_{i}$ denotes the number of data streams transmitted from BS $i$. For the sake of brevity, we assume that each BS serves only one user in the totally overlapped area ($\mathcal{K} = 1$), and the received signal at the user of interest can be written as, \cite{galym},
\begin{align}
	\label{Y1}
	\tilde{\mathbf{y}}_{j} = \underset{\textrm{desired + XCI signals}}{\underbrace{\mathbf{U}_{j}^{H}\sum\limits_{i=j-\mathcal{L}+1}^{j} \mathbf{H}_{j,i}\mathbf{V}_{i}\mathbf{s}_{i}}} + \underset{\textrm{ICI}}{\underbrace{\mathbf{U}_{j}^{H} \sum\limits_{l=1,l\not=i}^{L} \mathbf{H}_{j,l}\mathbf{V}_{l}\mathbf{s}_{l}}}+\tilde{\mathbf{n}}_{j},~\forall j\in L,
\end{align}
where $\mathbf{U}_{j}\in\mathbb{C}^{N_{j}\times d^{[j]}}$ denotes the receive suppression matrix of user $j$, and $\tilde{\mathbf{n}}_{j} = \mathbf{U}_{j}^{H} \mathbf{n}_{j} \in \mathbb{C}^{d^{[j]} \times 1}$ is the effective zero-mean additive white Gaussian noise (AWGN) vector, with $\mathbb{E}\left\lbrace \tilde{\mathbf{n}}_{j}\tilde{\mathbf{n}}_{j}^{H} \right\rbrace =\sigma_{\tilde{n}}^{2}\mathbf{I}$. $d^{[j]}$ is the number of data streams required to be decoded at user $j$. $\mathbf{H}_{j,i} \in \mathbb{C}^{N_{j} \times M_{i}}$ is the channel matrix between BS $i$ and user $j$. As in \cite{galym}, the transmit BF design can be proceeded as 
\begin{equation}
	\mathbf{V}_{i} = \mathbf{V}_{i}^{[ICI]} \times \mathbf{V}_{i}^{[XCI]},~\forall i\in L,
\end{equation}
where $\mathbf{V}_{i}^{[ICI]}\in\mathbb{C}^{M_{i} \times Q_{i}}$ and $\mathbf{V}_{i}^{[XCI]}\in\mathbb{C}^{Q_{i} \times d_{i}}$ are the BF matrices responsible for mitigating the ICI and XCI terms in \eqref{Y1}, respectively. $\mathbf{V}_{i} \in \mathbb{C}^{M_{i} \times d_{i}}$ is the transmit BF matrix at BS $i$, with $\textrm{tr}\left\lbrace \mathbf{V}_{i} \mathbf{V}_{i}^{H} \right\rbrace = 1$. We assume that $\mathbf{s}_{i} \in \mathbb{C}^{d \times 1}$ is the data vector comprising of the symbols drawn from independent and identically distributed (i.i.d.) Gaussian input signaling and chosen from a desired constellation, with $\mathbb{E}\left\lbrace \mathbf{s}_{i}\mathbf{s}_{i}^{H} \right\rbrace = \mathbf{I}$. Then, all these conditions sufficiently meet the average power constraint at the BS.
	
With this in mind and due to the nature of broadcast transmissions, we can assume that the users within one cell desire to receive the same data, thus the transmitted signal can be expressed as
\begin{multline}
\label{signal_Tx}
\mathbf{s}_{i}=\left[ \mathbf{c}^{[i,i]T} ~~~ \mathbf{c}^{[i,i+1]T} ~~~ \cdots ~~~ \mathbf{c}^{[i,l]T} ~~~ \cdots ~~~ \mathbf{c}^{[i,i+\mathcal{L}-1]T} \right]^{T},~\forall i\in L,
\end{multline}
where $\mathbf{c}^{[i,l]} \in \mathbb{C}^{d_{[i,l]} \times 1}$ and $d_{[i,l]}$ indicate the message vector and its corresponding number of data streams transmitted from BS $i$ to the user in cell $l$, respectively. More specifically, \eqref{signal_Tx} implies that BS $i$ dedicates the first part of the data to its corresponding user $i$, while the second part of the data is transmitted to the user belonging to the neighbouring cell $(i+1)$, and so on. The superscript numeration $l \in \{i,i+1,\ldots,i+\mathcal{L}-1\}$ changes circularly such that\footnote{It is worth to note that all notations with $^{*}(\cdot)$ should satisfy Eq. \eqref{mod}.}
\begin{equation}
\label{mod}
l = \begin{cases}
\begin{array}{ll}
L, & \textrm{if}~\textrm{mod}_{L}(l) = 0,\\
\textrm{mod}_{L}(l), & \textrm{otherwise}.
\end{array}
\end{cases}
\end{equation}
For instance, for $i=4$ and $L=4~(\mathcal{L}=3)$, we have $\textrm{mod}_{L}\left(\{i, i+1, i+2\}\right) = \textrm{mod}_{4}\left(\{4, 5, 6\}\right) \rightarrow \{4,1,2\}$. 

To decode the required signal successfully in \eqref{Y1}, the desired signal should be aligned into a subspace at the receiver side such that both ICI and XCI interfering signals are aligned into the subspace that is orthogonal to the desired signal. Therefore, the following conditions must be satisfied at user $j$
\begin{align}
\label{XCI_cond}
&\sum_{i = j - \mathcal{L} + 1}^{j} \textrm{rank}\left(\mathbf{U}^{H}_{j}\mathbf{H}_{j,i}\mathbf{V}_{i}\right)=d^{[j]},~\forall j\in L,\\
\label{ICI_cond}
&~~~\sum_{l = 1}^{L}\mathbf{U}^{H}_{j}\mathbf{H}_{j,l}\mathbf{V}_{l} = \mathbf{0},~\forall j\in L,~l\not=i,
\end{align}
where $d^{[j]}$ is the number of resolvable interference-free data streams at user $j$. Condition \eqref{XCI_cond} ensures that the user of interest can extract the desired signal coming from the channel links simultaneously carrying both the desired and XCI signals, while \eqref{ICI_cond} implies that the ICI signals are totally mitigated at the receiver side. 
	
\subsection{Imperfect CSI}
Since the CSI acquisition in practice is presented by imperfect estimates of the channel parameters, the system performance is likely to be degraded. Similar to the same assumption as in \cite{R22}, \cite{R23}, \cite{R24}, we assume that the precoders and combiners are designed with the knowledge of CSI mismatch. Similar to \cite{Razavi1}, \cite{Razavi2}, we exploit the following model to construct the CSI mismatch as
\begin{equation}
\hat{\mathbf{H}} = \mathbf{G} + \mathbf{E},
\end{equation} 
where the actual channel realization $\mathbf{G}$ is thought to be independent of the channel measurement error matrix $\mathbf{E}$. Defining the nominal SNR as $\rho = \frac{P}{\sigma_{n}^{2}}$, we further regard $\mathbf{E}$ as a Gaussian matrix where each entry is generated using i.i.d. random variables with zero mean and variance $\tau$ such that
\begin{equation}
\label{ErrorMatrix}
\textrm{vec}\left(\mathbf{E}\right) \sim \mathcal{CN}\left({\boldsymbol{0}}, \tau \mathbf{I} \right)~\textrm{with}~\tau \triangleq \beta \rho^{-\alpha},~\beta > 0,~\alpha \ge 0, 
\end{equation}
where $\textrm{vec}\left(\cdot\right)$ indicates the vector operator and $\mathcal{CN}(\boldsymbol{\mu},\mathbf{\Sigma})$ denotes the multivariate complex normal distribution with mean vector $\boldsymbol{\mu}$ and covariance matrix $\mathbf{\Sigma}$. 

In this model, the error variance can be used to capture a variety of CSI acquisition scenarios, i.e., dependent on the SNR $(\alpha\not=0)$ or be independent of that $(\alpha=0)$. In particular, perfect CSI scenario can be obtained by setting $\tau = 0$ $(\alpha \rightarrow \infty)$. Two distinct cases of CSI acquisition, known as reciprocal and CSI feedback channels, can be described by setting $\alpha = 0$ and $\alpha = 1$, respectively.

To facilitate further analysis, it is more appropriate to derive the statistical properties of $\mathbf{G}$ conditioned on $\hat{\mathbf{H}}$. Since $\hat{\mathbf{H}} = \mathbf{G} + \mathbf{E}$, with $\mathbf{G}$ and $\mathbf{E}$ being statistically independent Gaussian variables, $\mathbf{G}$ and $\hat{\mathbf{H}}$ are jointly Gaussian. Therefore, after conditioning on $\hat{\mathbf{H}}$, \cite{Paula}, \cite{Kay}, the real channel realization can be expressed as
\begin{equation}
	\label{Hr}
	\mathbf{G} = \frac{1}{1+\tau}\hat{\mathbf{H}} + \hat{\mathbf{E}},
\end{equation}
where $\textrm{vec}\left( \hat{\mathbf{E}} \right) \sim \mathcal{CN}\left( \boldsymbol{0}, \frac{\tau}{1+\tau}\mathbf{I} \right)$ is statistically independent of $\hat{\mathbf{H}}$.
	
\subsection{Kronecker Product based Channel Modeling with Antenna Correlation}
The downlink channel $\mathbf{H}_{j,i}$ in \eqref{Y1} is modeled as a correlated flat fading channel. Since the system model in \cite{galym} presumes that both transmitter (Tx) and receiver (Rx) nodes are deployed with multiple antennas, we assume that the fading is correlated at both sides. 
	
The general model of the correlated channel is given as, \cite{kotecha}, \cite{raghavan}, \cite{love},
\begin{equation}
	\textrm{vec}\left(\mathbf{H}\right)=\mathbf{R}^{1/2}\textrm{vec}\left(\mathbf{G}\right),
\end{equation}
where $\mathbf{G}$ is the i.i.d. MIMO channel with either perfect CSI or CSI mismatch as in \eqref{Hr}, and $\mathbf{R}$ is the covariance matrix defined as 
\begin{equation}
	\mathbf{R}\triangleq\mathbb{E}\left\lbrace \textrm{vec}\left(\mathbf{G}\right)\textrm{vec}\left(\mathbf{G}\right)^{H} \right\rbrace.
\end{equation}
	
A popular MIMO channel simulation model is the Kronecker model, \cite{love}, \cite{validity}, \cite{loyka}. The Kronecker model is limited because it does not take into account the coupling between the direction of departure (DoD) at the transmitter and the direction of arrival (DoA) at the receiver, which is typical for MIMO channels. Despite the limitation of ignoring the coupling between the DoD and DoA at the transmit and receive ends, the Kronecker model is widely used in information theoretic capacity analysis and simulation studies, \cite{Kron1}, \cite{Kron2}, \cite{Kron3}, \cite{Kron4}. Though the Kronecker model is expected to be inaccurate for increasing array size and angular resolution, it still finds use in large MIMO system studies because of its simplicity. Therefore, the Kronecker model is found to be sufficient to analyse the effect of spatial correlation on the network performances.

Since this model is based on the assumption of separability of the transmit and receive correlation coefficients, the transmitter does not affect the spatial properties of the received signal. In such a case the correlation matrix is given as
\begin{equation}
	\mathbf{R}=\mathbf{R}_{r}\otimes\mathbf{R}_{t},
\end{equation} 
where the notation $\otimes$ stands for the Kronecker product operator, $\mathbf{R}_{r}\in\mathbb{C}^{N\times N}$ and $\mathbf{R}_{t}\in\mathbb{C}^{M\times M}$ are the receive and transmit correlation matrices defined as 
\begin{align}
& \mathbf{R}_{r}=\frac{1}{M}\mathbb{E}\left\lbrace \mathbf{G}\mathbf{G}^{H} \right\rbrace,\\
& \mathbf{R}_{t}=\frac{1}{N}\mathbb{E}\left\lbrace \mathbf{G}^{H}\mathbf{G} \right\rbrace.
\end{align}

Then, the channel realization, with respect to imperfect CSI \eqref{Hr}, can be modeled as, \cite{gesbert}, \cite{chris},
\begin{align}
	\label{Hcorr}
	\mathbf{H}&=\mathbf{R}_{r}^{1/2}\mathbf{G}\mathbf{R}_{t}^{1/2}=\mathbf{R}_{r}^{1/2}\left(\frac{1}{1+\tau} \hat{\mathbf{H}} + \hat{\mathbf{E}}\right)\mathbf{R}_{t}^{1/2}\nonumber\\
	&=\frac{1}{1+\tau} \mathbf{R}_{r}^{1/2}\hat{\mathbf{H}}\mathbf{R}_{t}^{1/2} + \mathbf{R}_{r}^{1/2}\hat{\mathbf{E}}\mathbf{R}_{t}^{1/2} =\tilde{\mathbf{H}} + \tilde{\mathbf{E}},
\end{align}
where $\tilde{\mathbf{E}} = \mathbf{R}_{r}^{1/2}\hat{\mathbf{E}}\mathbf{R}_{t}^{1/2}\sim\mathcal{CN}(\mathbf{0},\mathbf{R}_{r} \otimes \frac{\tau}{\tau+1}\mathbf{I} \otimes \mathbf{R}_{t})$ is the estimation error which is uncorrelated with $\tilde{\mathbf{H}} = \frac{1}{1+\tau} \mathbf{R}_{r}^{1/2}\hat{\mathbf{H}}\mathbf{R}_{t}^{1/2}$. 
	
With this in mind, we refer to \cite{chizhik}, where, for the case of both-end correlation, the authors showed that the Kronecker-based exponential and uniform models are empirically reasonable to apply at the transmitter and receiver sides, respectively. Moreover, these simple single-parameter models allow one to investigate the effects of both-end correlation on the achievable sum rate and DoF to achieve clearer insights in an explicit way.
	
\subsubsection{Exponential correlation coefficient model}
The most common and easy model that accounts for potential antenna correlation is the exponential model, \cite{love}, \cite{loyka}, which can be accordingly utilized at the transmitter side as follows
\begin{equation}
	\label{corrExp}
	\mathbf{R}_{t_{[m,n]}}=\begin{cases}
	\begin{array}{ll}
	r^{|m-n|}, &\textrm{if} ~ m\geq n, \\
	(r^{\dagger})^{|m-n|}, &\textrm{if} ~ m < n,
	\end{array}
	\end{cases}
\end{equation}
where the subscript notation $[m,n]$ indicates the matrix element located in the $m^{th}$ row and $n^{th}$ column. $(^{\dagger})$ denotes a complex conjugate and $r=ae^{j\theta}$ is the complex correlation coefficient with $0\leq a<1$. For simplicity, we assume that $r=a$ throughout the paper, unless it is restated. 
	
\subsubsection{Uniform correlation coefficient model}
The uniform coefficient model, \cite{salous}, is the worst case scenario where the correlation coefficients are defined as
\begin{equation}
	\label{corrUni}
	\mathbf{R}_{r_{[m,n]}}=\begin{cases}
	\begin{array}{ll}
	r^{|m-n|}, &\textrm{if} ~ m=n,\\
	r, &\textrm{if} ~ m\not=n,
	\end{array}
	\end{cases}
\end{equation}
where we assume that the coefficients of all neighboring subchannels are equal to those of the distant channels.

\section{The IA Feasibility Conditions and DoF Region Analysis}
As stated in \cite{galym}, \cite{galym1}, the compounded MIMO BC scenario represents the network model where the users located in the totally overlapped area experience a multi-source transmission by $\mathcal{L}$ BSs. In particular, the authors in \cite{galym} focused on the scenario when each Tx-Rx pair has an identical antenna configuration with an equal number of data streams transmitted to the users of interest. In contrast to \cite{galym}, in this work, we consider the network scenario when the BSs serve mixed user classes. Therefore, we exploit a three-cell network scenario to analyze how the proposed IA scheme performs under this transmission strategy. For the sake of simplicity, we assume that each cell consists of one BS serving a single user residing in the totally overlapped area. Thus, each BS aims to send data to two users of interest ($\mathcal{L}=2$). Accordingly, the transmitted signal in \eqref{signal_Tx} can be rewritten as
\begin{align}
	\label{signal_Tx1}
	\mathbf{s}_{i}=\left[ \mathbf{c}^{[i,i]T} ~~~ \mathbf{c}^{[i,i+1]T} \right]^{T},~\forall i\in L,
\end{align}
where $\mathbf{c}^{[i,i]} \in \mathbb{C}^{d_{[i,i]} \times 1}$ is the data vector transmitted from BS $i$ to the corresponding user belonging the same cell, while $\mathbf{c}^{[i,i+1]} \in \mathbb{C}^{d_{[i,i+1]} \times 1}$ indicates the data transmitted to a desired user in the neighboring cell. 
	
\begin{figure}[h]
	\centering
	\includegraphics[width=0.5\columnwidth]{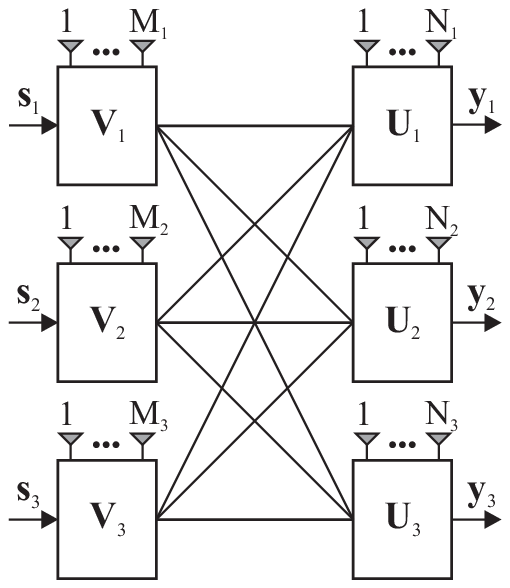}
	\caption{A system model with three-cell compounded MIMO BCs for $\mathcal{K} = 1$ user per cell with various numbers of antennas at each Tx-Rx pair.}
	\label{3us_refer}
\end{figure}

In the following sections, we show how to define the antenna configuration for each Tx-Rx pair to achieve a given DoF in the three-cell MIMO network (see Fig. \ref{3us_refer}) and determine the performance metrics accounting for the impact of spatial correlation.
	
\subsection{The Feasibility Conditions of Interference Alignment}
Since we decouple the transmit BF matrix into two parts, we need to start with the design of the part responsible for IC interference mitigation. For the sake of brevity, we refer to \cite{galym}, where the given scheme is also applicable for the case with a single user per cell. 
	
After applying that scheme, we come up with the $\mathbf{V}_{i}^{[ICI]}$ matrix with the dimension of $M_{i} \times (M_{i} - N_{i-1})$. Hence, the matrix dimension leads to the following condition
\begin{equation}
	\label{difference}
	(M_{i}-N_{i-1})^{+} \ge d_{i},~\forall i\in L
\end{equation}
to be satisfied for a successful transmission of all the desired data to the interested users, and $(\cdot)^{+}$ represents a function that returns a non-negative value. $d_{i}$ is the number of data streams transmitted from BS $i$ and defined with respect to \eqref{signal_Tx1} as
\begin{equation}
	d_{i} = \sum\limits_{j=i}^{i+1}d_{[i,j]},~\forall i\in L.
\end{equation}
For simplicity of the following derivations, we introduce a new variable indicating the number of columns in $\mathbf{V}_{i}^{[ICI]}$, as well as the number of non-occupied antennas at the transmitter side, and define
\begin{equation}
	\label{Qi}
	Q_{i} = (M_{i} - N_{i-1})^{+},~\forall i\in L.
\end{equation}
Finally, we restate the condition of the successive ICI cancellation given in \eqref{difference} as
\begin{equation}
	\textrm{rank}\left(\mathbf{V}_{i}^{[ICI]}\right)\ge Q_{i},~\forall i\in L.
\end{equation} 
	
Since we have different classes of receivers and BSs that aim to transmit different numbers of data streams to the users of interest, we need to define the number of data streams desired to be decoded at user $j$ as 
\begin{equation}
	d^{[j]}=\sum\limits_{\forall i}^{}d_{[i,j]},~\forall j\in L.
\end{equation} 
In general, the number of transmitted data streams from BS $i$ is not equal to the number of data streams desired to be received by the corresponding user $i$, $d_{i}\not=d^{[i]}$; nevertheless, the total number of streams at the transmitter side always matches the one at the receiver side as
\begin{equation}
	\sum\limits_{i=1}^{L=3}d_{i}=\sum\limits_{j=1}^{L=3}d^{[j]}.
\end{equation}
	
With this in mind, the received signal at the user of interest can be written as
\begin{align}
	\label{Y2}
	\tilde{\mathbf{y}}_{j}&=\underset{\textrm{desired + XCI signals}}{\underbrace{\mathbf{U}_{j}^{H}\sum\limits_{i=j-1}^{j} \mathbf{H}_{j,i}\mathbf{V}_{i}^{[ICI]}\mathbf{V}_{i}^{[XCI]}\mathbf{s}_{i}}} + \underset{\textrm{ICI=0}}{\underbrace{\cancel{\mathbf{U}_{j}^{H} \sum\limits_{l=1,l\not=i}^{L=3} \mathbf{H}_{j,l}\mathbf{V}_{l}^{[ICI]}\mathbf{V}_{l}^{[XCI]}\mathbf{s}_{l}}}}+\tilde{\mathbf{n}}_{j}\nonumber\\
	&=\sum\limits_{i=j-1}^{j}\mathbf{U}_{j}^{H}\mathbf{H}_{j,i}\mathbf{V}_{i}^{[ICI]}\mathbf{V}_{i}^{[XCI]}\mathbf{s}_{i}+\tilde{\mathbf{n}}_{j} =\sum\limits_{i=j-1}^{j}\bar{\mathbf{H}}_{j,i}\mathbf{V}_{i}^{[XCI]}\mathbf{s}_{i}+\tilde{\mathbf{n}}_{j}\nonumber\\
	&=\sum\limits_{i=j-1}^{j}\bar{\mathbf{H}}_{j,i}  \underset{Q_{i}\times d_{[i,i]}}{\underbrace{\left[\mathbf{V}^{[i,i]XCI}\right.}} ~~~ \underset{Q_{i}\times d_{[i,i+1]}}{\underbrace{\left.\mathbf{V}^{[i,i+1]XCI}\right]}} \left[\begin{array}{c}
	\mathbf{c}^{[i,i]} \\
	\mathbf{c}^{[i,i+1]}
	\end{array}\right] + \tilde{\mathbf{n}}_{j},~\forall j\in L,
\end{align}
where $\bar{\mathbf{H}}_{j,i} = \mathbf{U}_{j}^{H}\mathbf{H}_{j,i}\mathbf{V}_{i}^{[ICI]}$ indicates the effective channel matrix. The under-braced terms represent the $Q_{i}\times d_{[m,n]}$ matrices responsible for the XCI cancellation of the undesired data, $\mathbf{c}^{[m,n]}$. Then, the user of interest no longer experiences ICI, but the interference arriving along the desired directions presented by the $\mathbf{c}^{[j-1,j-1]}$ and $\mathbf{c}^{[j,j+1]}$ data vectors with $d_{[j-1,j-1]}$ and $d_{[j,j+1]}$ data streams, respectively. Hence, we define the minimum and maximum numbers of the interfering data streams as 
\begin{align}
	\label{k_i}
	k_{i}&=\min\left\lbrace d_{[j-1,j-1]},d_{[j,j+1]} \right\rbrace,\\
	\label{r_i}
	r_{i}&=\max\left\lbrace d_{[j-1,j-1]},d_{[j,j+1]} \right\rbrace, 
\end{align}
and the corresponding difference as
\begin{equation}
	\label{wi}
	w_{i}=r_{i}-k_{i}.
\end{equation}
	
To be more specific, we consider the received signal at user 1, which can be presented with respect to \eqref{signal_Tx1} as
\begin{align}
	\label{Yr1}
	\tilde{\mathbf{y}}_{1}&=\bar{\mathbf{H}}_{1,1}\left[ \mathbf{V}^{[1,1]XCI}~~\mathbf{V}^{[1,2]XCI} \right] \left[\begin{array}{c}
	\mathbf{c}^{[1,1]}\\
	\mathbf{c}^{[1,2]}
	\end{array}\right] + \bar{\mathbf{H}}_{1,3}\left[ \mathbf{V}^{[3,3]XCI}~~\mathbf{V}^{[3,1]XCI} \right] \left[\begin{array}{c}
	\mathbf{c}^{[3,3]}\\
	\mathbf{c}^{[3,1]}
	\end{array}\right]
	+ \tilde{\mathbf{n}}_{1} \nonumber \\
	&=\underset{\textrm{desired signal}}{\underbrace{\bar{\mathbf{H}}_{1,1}\mathbf{V}^{[1,1]XCI}\mathbf{c}^{[1,1]} + \bar{\mathbf{H}}_{1,3}\mathbf{V}^{[3,1]XCI}\mathbf{c}^{[3,1]}}}  +\tilde{\mathbf{n}}_{1} + \underset{\textrm{interference}}{\underbrace{\bar{\mathbf{H}}_{1,1}\mathbf{V}^{[1,2]XCI}\mathbf{c}^{[1,2]} + \bar{\mathbf{H}}_{1,3}\mathbf{V}^{[3,3]XCI}\mathbf{c}^{[3,3]}}}.
\end{align}

Since we assume that user 1 is deployed with $N_{1}$ antennas, this number has to be enough to allocate the desired signal into a subspace separate from the interference, and can be then defined as 
\begin{equation}
	\label{N1}
	N_{1}\ge d^{[1]}+d_{I^{[1]}},
\end{equation} 
where $d_{I^{[1]}}$ is the subspace spanning the interfering $\mathbf{c}^{[1,2]}$ and $\mathbf{c}^{[3,3]}$ vectors present at user 1 and can be expressed as
\begin{equation}
	\label{I1}
	I^{[1]}=\sum\limits_{i=1}^{d_{[1,2]}}\bar{\mathbf{H}}_{1,1} \mathbf{v}_{i}^{[1,2]}c_{i}^{[1,2]}
	+\sum\limits_{i=1}^{d_{[3,3]}}\bar{\mathbf{H}}_{1,3} \mathbf{v}_{i}^{[3,3]}c_{i}^{[3,3]}.
\end{equation}
Therefore, the number of receive antennas has to be enough to decode the received signal and consequently needs to satisfy the following requirement
\begin{equation*}
	N_{1}\ge d^{[1]}+d_{[1,2]}+d_{[3,3]},
\end{equation*}
which is not always possible to provide at the receive side due to the physical space limitation. 
	
We define the $k_{1}$ and $r_{1}$ variables given in \eqref{k_i}\textendash\eqref{r_i} as
\begin{align*}
	k_{1}& = \min\left( d_{[1,2]},d_{[3,3]} \right),\\
	r_{1}& = \max\left( d_{[1,2]},d_{[3,3]} \right).
\end{align*}
Accordingly, the corresponding difference between $d_{[1,2]}$ and $d_{[3,3]}$ is given as 
\begin{equation*}
	w_{1} = r_{1} - k_{1},
\end{equation*}
where $w_{1}$ indicates the number of the interfering data streams that can not be aligned with the other interfering signal, and, therefore, these $w_{1}$ vectors need to be mitigated at the receiver.
	
To reduce the subspace spanning the $I^{[1]}$, we make sure that these interfering signals span a one-dimensional space, and determine the $k_{1}$ pairs of precoding vectors such that
\begin{align}
	\label{equal1}
	\bar{\mathbf{H}}_{1,1}\mathbf{v}_{i}^{[1,2]}=-\bar{\mathbf{H}}_{1,3}\mathbf{v}_{i}^{[3,3]},~\forall i\in\{1,2,\ldots,k_{1}\},
\end{align}
where $\mathbf{v}_{i}^{[m,n]}$ is the $i^{th}$ column of the $\mathbf{V}^{[m,n]XCI}$ matrix from \eqref{Yr1}, and the $XCI$ notation is omitted for brevity. We randomly pick the $\mathbf{v}^{[1,2]}_{1,\ldots,k_{1}}$ and $\mathbf{v}^{[3,3]}_{1,\ldots,k_{1}}$ vectors to guarantee that they are linearly independent with probability one. This can be presented as shown in \eqref{I2}, where we assumed that $r_{1} = d_{[1,2]}$. 
	
\begin{figure*}[!t]
	\normalsize
	\begin{align}
	\label{I2}
	I^{[1]}&=\sum\limits_{i=1}^{d_{1,2}}\bar{\mathbf{H}}_{1,1} \mathbf{v}_{i}^{[1,2]}c_{i}^{[1,2]}
	+\sum\limits_{i=1}^{d_{3,3}}\bar{\mathbf{H}}_{1,3} \mathbf{v}_{i}^{[3,3]}c_{i}^{[3,3]} \nonumber\\
	&=\underset{\textrm{space dimension range = 1}}{\underbrace{\sum\limits_{i=1}^{k_{1}}\left( \bar{\mathbf{H}}_{1,1}\mathbf{v}_{i}^{[1,2]}c_{i}^{[1,2]} + \bar{\mathbf{H}}_{1,3}\mathbf{v}_{i}^{[3,3]}c_{i}^{[3,3]} \right)}}
	+ \underset{\textrm{space dimension range = 0}}{\underbrace{\sum\limits_{l=k_{1}+1}^{k_{1}+w_{1}}\bar{\mathbf{H}}_{1,1} \mathbf{v}_{l}^{[1,2]}c_{l}^{[1,2]}}}
	\end{align}
	\hrulefill
\end{figure*}
	
Since user 1 obtains the $\mathbf{c}^{[1,2]}$ and $\mathbf{c}^{[3,3]}$ data vectors with $d_{[1,2]}$ and $d_{[3,3]}$ data streams, respectively, for $d_{1,2}\not=d_{[3,3]}$, we then need to define which effective channel matrix, $\bar{\mathbf{H}}_{1,x}$, needs to be cancelled, where $x$ can be derived from
\begin{equation*}
	\{x,y\} = \left\lbrace 
	\begin{alignedat}{2}
	& \{1,2\},&~\textrm{if}~r_{1}=d_{[1,2]},\\
	& \{3,3\},&~\textrm{if}~r_{1}=d_{[3,3]}.
	\end{alignedat}\right. 
\end{equation*}
Therefore, we have the $w_{1}$ interfering BF vectors, $\mathbf{V}^{[x,y]}_{\{k_{1}+1:k_{1}+w_{1}\}} = \left\lbrace \mathbf{v}_{k_{1}+1}^{[x,y]},\ldots,\mathbf{v}_{k_{1}+w_{1}}^{[x,y]} \right\rbrace$,  that are obtained by finding the null space of the corresponding effective channel as follows
\begin{equation}
	\mathbf{V}^{[x,y]}_{\{k_{1}+1:k_{1}+w_{1}\}}=\textrm{null}\left( \bar{\mathbf{H}}_{1,x} \right),
\end{equation}
where $\mathbf{V}_{\{k_{1}+1:k_{1}+w_{1}\}}^{[x,y]}$ denotes the part of the BF matrix with a range from the $(k_{1}+1)^{th}$ up to the $(k_{1}+w_{1})^{th}$ column. This leads to the following condition to be satisfied 
\begin{equation}
	\label{Q_w}
	w_{1} \le \left(Q_{1}-N_{1}\right)^{+}~\textrm{or}~w_{1} \le \left(Q_{3}-N_{1}\right)^{+}.
\end{equation}
As a result, the interference observed by user 1 can be mitigated as shown in \eqref{inter1}, where we assume $d_{[1,2]}>d_{[3,3]}$ and the boxed term is redundant, unless $d_{[1,2]}<d_{[3,3]}$. According to \eqref{I2}, the interference at the user of interest spans a one-dimensional space, thus we redefine the number of receive antennas given in \eqref{N1} as
\begin{equation}
	N_{1}=d^{[1]}+1.
\end{equation}
Finally, if all the conditions above are satisfied, user 1 experiences the interference-free data transmission.

Similar to cell 1, we define the conditions satisfying the ability to maintain the proposed scheme for the rest of cells.	
\begin{figure*}[!t]
	\normalsize
	\begin{equation}
		\label{inter1}
		\underset{N_{1}\times(Q_{1}+Q_{3})}{\underbrace{\left[\bar{\mathbf{H}}_{1,1}~~\bar{\mathbf{H}}_{1,3}\right]}} 
		\underset{(Q_{1}+Q_{3})\times (k_{1}+w_{1})}{\underbrace{\left[ \begin{array}{ccc|ccc}
		| & \cdots & | & | & \cdots & | \\
		\mathbf{v}_{1}^{[1,2]} & \cdots & \mathbf{v}_{k_{1}}^{[1,2]} & \mathbf{v}_{k_{1}+1}^{[1,2]} & \cdots & \mathbf{v}_{k_{1}+w_{1}}^{[1,2]} \\
		| & \cdots & | & | & \cdots & |  \\
		\hline
		| & \cdots & | & | & \cdots & | \\
		\mathbf{v}_{1}^{[3,3]} & \cdots & \mathbf{v}_{k_{1}}^{[3,3]} & \mathbf{0} & \cdots & \mathbf{0} \\
		| & \cdots & | & | & \cdots & | 
		\end{array}\right.
		\left.\begin{array}{|ccc|}
		\hline
		| & \cdots & | \\ 
		\mathbf{0} & \cdots & \mathbf{0} \\
		| & \cdots & | \\ \hline
		| & \cdots & | \\
		\mathbf{v}_{k_{1}+1}^{[3,3]} & \cdots & \mathbf{v}_{k_{1}+w_{1}}^{[3,3]}\\
		| & \cdots & | \\
		\hline
		\end{array}
		~\right]}}
		= 
		\underset{N_{1}\times \max(d_{[1,2]},d_{[3,3]})}{\underbrace{\left[\begin{array}{ccc}
		| & \cdots & | \\
		\mathbf{0} & \cdots & \mathbf{0}\\
		| & \cdots & |
		\end{array}\right]}}
		\end{equation}
		\hrulefill
\end{figure*}

Below we briefly explain how to derive the receive suppression matrices. The given example will be generalized for user $i$. Due to the help of transmit beamforming matrices, we ensure that the interfering signals coming from BSs $j~ (\forall j\in L=3, j\not=i)$ are aligned along one shared space (for brevity, we denote this effective interference as $\tilde{\mathbf{H}}_i$). Having this, we define the space spanning the interference as
\begin{equation}
\mathbf{T}_i = \textrm{span}\{\tilde{\mathbf{H}}_i^{H}\mathbf{U}_i\},
\end{equation}
where $\textrm{span}\{\cdot\}$ denotes the subspace spanned by the column vectors of a matrix. Next, we rewrite this equation as
\begin{equation}
\left[\mathbf{I}_{M_i} ~~~ -\tilde{\mathbf{H}}_i^{H} \right] \left[\begin{array}{c}
\mathbf{T}_i\\
\mathbf{U}_i
\end{array}\right] = \mathbf{F}_i \mathbf{W}_i = \mathbf{0}.
\end{equation}
Since the size of the matrix $\mathbf{F}_i$ is $M_i \times M_i + P_i$ ($P_i = \max(d_j, d_k)$ and $d_j$ and $d_k$ is the numbers of data streams transmitted from BSs $j$ and $k$, where $i\not=\{j,k\}$), the null space always exists because $M_i + P_i > M_i, \forall P_i > 0$. For more details, the reader can refer to \cite{jie}.

\subsection{The DoF Region Analysis}
With respect to the number of cells $L$, we define an upper bound on the DoF region. Thus, for the compounded MIMO BC network scenario with $L$ cells, we have $((\mathcal{L}+2)L+1)$ constraints that determine the upper limit of the DoF region.

While the set $\mathcal{D}_{out}$ describes an outer limit for all achievable $d_{[i,j]}$ on the compounded $L-$cell MIMO BC channel, maximization of the sum of $d_{[i,j]}$ over $\mathcal{D}_{out}$ is a linear LP problem, where it is required to explicitly estimate all extreme points of the feasible space, to calculate the objective values and, finally, to eliminate all redundant limits.
\begin{subequations}
	\begin{align}
	\mathcal{D}_{out}\triangleq \textrm{maximize}\quad &  \sum_{i = 1}^{L} \sum_{j = i}^{^{*}\textrm{mod}_L(i+(L - \mathcal{L}))} d_{[i,j]} \in \mathbb{R}^{L\mathcal{L}}_{+}, \nonumber \\
	\textrm{subject to}\quad & ~~ \hspace{-0.3cm}
	d_i = \sum_{j=i}^{^{*}\textrm{mod}_L(i + (L - \mathcal{L}))}
	d_{[i,j]} \le Q_{i}, \label{eq:subeq1} \\
	& ~~ \underset{d^{[i]}}{\underbrace{
			\sum_{j = i}^{^{*}\textrm{mod}_L(i+(L - 1))} d_{[j,i]}}} + 1 \le N_{i}, \label{eq:subeq2} \\
	& ~ \hspace{-0.3cm} D_{i}^{[k]} = d^{[j]} \cup Q_i \le \max(N_{j},Q_{i}), \label{eq:subeq7} \\
	& ~~j\in \underset{\textrm{$k$~samples}}{\underbrace{\{i,\ldots,^{*}\textrm{mod}_L(i + (L - \mathcal{L}))\}}},  \nonumber\\
	& ~~ \forall i\in L, k\in\{1,\ldots,(L - \mathcal{L})\}, \nonumber\\
	& ~~ \overline{Q}_{t} = w_{j} + N_{j},~\forall j\in L, \label{eq:subeq13} \\
	\textrm{where}\quad & ~~ t=m \leftarrow d_{[m,n]} = r_{j}~\textrm{in}~ \eqref{r_i}, \label{eq:subeq14}
	\end{align}
\end{subequations}
where $t$ and $\cup$ indicate the first subscript index and the union operation, respectively. The case of $w_{j} = 0$ implies that at this particular user the numbers of interfering streams are identical. $\overline{Q}$ indicates the variable that captures uncertainty of which $Q$ should be chosen according to \eqref{Q_w}.
		
Although at a glance it might seem that the DoF region above is not too different from the one presented in \cite{galym1}, it is worth mentioning that \cite{galym1} focused on the network case with an identical antenna configuration leading to identical numbers of data streams causing the interference at the receiver side, that is, in \cite{galym1} the authors considered the users of one user class. Moreover, the algorithm in \cite{galym1} can not be applied for the compounded MIMO BC with mixed user classes. Thus, considering different classes of users entails more conditions related to the XCI mitigation shown by the second term in \eqref{I2}, which spans a zero-dimensional space. 
		
The set $\mathcal{D}_{out}$ provides the conditions that define the outer limit for all the attainable $d_{[j,i]}$ under the compounded MIMO BC by maximizing a weighted sum of $d_{[j,i]}$ which is a linear programming problem. At the same time, this set of conditions is suitable either to analyse the achievable DoF region under a given network antenna configuration or to calculate the minimum number of antennas at each node in order to obtain the required DoF. Subsequently, we provide the algorithm that allows us to compute the minimum required number of antennas at each Tx-Rx pair with maximum DoF.

\begin{algorithm}                     
	\caption{Defining the Antenna Configuration}          
	\label{findme}                          
	\begin{algorithmic}[1]                   
		\REQUIRE  \\
		\textbf{inputs} $\left\lbrace d_{[i,j]} \right\rbrace,~\forall i,j \in L $ \\
		$Q_{i}$ and $N_{j}$ from \eqref{eq:subeq1}\textendash\eqref{eq:subeq2} 
		\ENSURE 
		\FOR{$j = 1:L$}
		\IF{\eqref{eq:subeq7} is not valid} 
		\STATE \textbf{update} $Q_i \gets D_i^{[k]}$ 
		\ENDIF
		\STATE \textbf{return} $Q_i$.
		\ENDFOR
		\FOR{$j = 1:L$}
		\STATE \textbf{calculate} $w_{j},$ according to \eqref{wi}.
		\STATE \textbf{find} the value of $t$ with respect to \eqref{eq:subeq14}.
		\IF{$t = j$}
		\STATE $\overline{Q}_{t} := \overline{Q}_{j}$
		\IF{$\overline{Q}_{j} > Q_{j}$} 
		\STATE \textbf{update} $Q_{j} \gets \overline{Q}_{j}$
		\ENDIF
		\ELSE
		\STATE $\overline{Q}_{t} := \overline{Q}_{j-1}$ 
		\IF{ $\overline{Q}_{j-1} > Q_{j-1}$}
		\STATE \textbf{update} $Q_{j-1} \gets \overline{Q}_{j-1}$ 
		\ENDIF
		\ENDIF
		\STATE \textbf{return} $Q_{j-1}$ or $Q_{j}$ (if updated only). 
		\STATE \textit{This derived value will be used in the next iteration}
		\ENDFOR
		\STATE \textbf{utilize} $Q_i$ and $N_j,~\forall i,j\in L$, to calculate the numbers of transmit antennas as in \eqref{Qi}.
	\end{algorithmic}
\end{algorithm}
		
The following provides the total number of DoF by explicitly solving the LP problem for the three-cell compounded MIMO BC case. 
{\allowdisplaybreaks
\begin{align}
	& Proposition~1: \nonumber \\
	\eta&\triangleq\underset{\mathcal{D}_{out}}{\max} 
	\sum_{i = 1}^{L=3} \sum_{j = i}^{^{*}\textrm{mod}_L(i+1)} d_{[i,j]} \\
	&=\min\left\lbrace Q_{1}+Q_{2}+Q_{3}, N_{1}+N_{2}+N_{3}-3,\right. \max(N_{1},Q_{1}) + \max(N_{3},Q_{2}), \nonumber \\
	& \max(N_{1},Q_{3}) + \max(N_{2},Q_{2}),~\max(N_{2},Q_{1}) + \max(N_{3},Q_{3}), \nonumber \\
	& \frac{\max(N_{1},Q_{1}) + \max(N_{1},Q_{3}) + \max(N_{3},Q_{2}) + \max(N_{2},Q_{2})}{2},\nonumber \\
	& \frac{\max(N_{1},Q_{3})+\max(N_{2},Q_{1}) + \max(N_{3},Q_{3})+\max(N_{2},Q_{2})}{2},\nonumber \\
	& \frac{\max(N_{1},Q_{1}) + \max(N_{2},Q_{1}) + \max(N_{3},Q_{2})+\max(N_{3},Q_{3})}{2}, \nonumber \\
	&\frac{\max(N_{1},Q_{1}) + \max(N_{1},Q_{3}) + Q_{2} + N_{2} + N_{3}}{2} - 1, \nonumber \\
	& \frac{\max(N_{1},Q_{1}) + \max(N_{2},Q_{1}) + Q_{2} + Q_{3} + N_{3} - 1}{2}, \nonumber \\
	& \frac{\max(N_{1},Q_{1}) + \max(N_{3},Q_{2}) + Q_{1} + Q_{2} + Q_{3}}{2}, \nonumber \\
	& \frac{\max(N_{1},Q_{1}) + \max(N_{3},Q_{2})+ N_{1} + N_{2} + N_{3} - 1}{2} - 1, \nonumber \\
	& \frac{\max(N_{1},Q_{3}) + \max(N_{2},Q_{2}) + Q_{1} + Q_{2} + Q_{3}}{2}, \nonumber \\
	& \frac{\max(N_{1},Q_{3}) + \max(N_{2},Q_{2}) + N_{1} + N_{2} + N_{3} - 1}{2} - 1, \nonumber \\
	& \frac{\max(N_{1},Q_{3}) + \max(N_{3},Q_{3}) + Q_{1} + Q_{2} + N_{2} - 1}{2}, \nonumber \\
	& \frac{\max(N_{2},Q_{1}) + \max(N_{2},Q_{2}) + Q_{3} + N_{1} + N_{3}}{2} - 1, \nonumber \\
	& \frac{\max(N_{2},Q_{1}) + \max(N_{3},Q_{3}) + Q_{1} + Q_{2} + Q_{3}}{2}, \nonumber \\
	& \frac{\max(N_{2},Q_{1}) + \max(N_{3},Q_{3}) + N_{1} + N_{2} + N_{3} - 1}{2} - 1,\nonumber \\
	& \frac{\max(N_{2},Q_{2}) + \max(N_{3},Q_{2}) + Q_{1} + Q_{3} + N_{1} - 1}{2}, \nonumber \\
	& \frac{\max(N_{3},Q_{2}) + \max(N_{3},Q_{3}) + Q_{1} + N_{1} + N_{2}}{2} - 1,\nonumber \\
	& \frac{\max(N_{1},Q_{1}) + \max(N_{1},Q_{3}) + \max(N_{2},Q_{1})}{3} \nonumber \\
	& \left.+ \frac{\max(N_{2},Q_{2}) + \max(N_{3},Q_{2}) + \max(N_{3},Q_{3})}{3}\right\rbrace. \nonumber 
\end{align}}
$\quad \textbf{Outline of the Proof}:$ Proposition 1 can be verified by solving the dual problem by linear programming $$\max\left( d_{[1,1]} + d_{[1,2]} + d_{[2,2]} + d_{[2,3]} + d_{[3,3]} + d_{[3,1]} \right).$$
Since all the extreme points of the feasible space can be directly evaluated, we compute the objective value at these points and eliminate the limits that can be regarded redundant. Using the fundamental theorem of LP, \cite{FundTheorem}, \cite{Luenberg}, we find the solution. For the sake of brevity, the derivation details are omitted. 
		
It is worth noting that all the terms given in Proposition 1 are essential because any of them is valid for a certain antenna configuration.
		
We define the achievable DoF for our multi-cell network as the pre-log factor of the sum rate, \cite{Gou}, \cite{ach_C}. This is one of the key metrics used for assessing the performance of a multiple antenna based system in the high SNR region defined as
\begin{equation}
	\label{eta}
	\eta = \lim\limits_{SNR\rightarrow\infty}\frac{\mathcal{I}_{\sum}(SNR)}{\log_{2}(SNR)} = \sum\limits_{j=1}^{L}d^{[j]},
\end{equation}
where $\mathcal{I}_{\sum}(SNR)$ denotes the sum rate that can be achieved at a given SNR defined as $\mathcal{I}_{\sum}(SNR) = \sum\limits_{j=1}^{L}\mathcal{I}_j$, where $\mathcal{I}_j$ and $d^{[j]}$ are the data rate and the number of the successfully decoded data streams at user $j$, respectively.
		
Therefore, the received signal at user $j$, with respect to the considered channel model in \eqref{Hcorr}, can be rewritten as
\begin{align}
	\label{Y3}
	\tilde{\mathbf{y}}_{j} &= \mathbf{U}_{j}^{H} \sum\limits_{i=1}^{L=3}\mathbf{H}_{j,i}\mathbf{V}_{i} \mathbf{s}_{i} + \tilde{\mathbf{n}}_{j} \nonumber\\
	&= \underset{\textrm{desired signal + XCI}}{\underbrace{\mathbf{U}_{j}^{H}\sum\limits_{i=1,~i\not=j+1}^{L=3}\mathbf{H}_{j,i}\mathbf{V}_{i} \mathbf{s}_{i}}} + 
	\underset{\textrm{ICI}}{\underbrace{\mathbf{U}_{j}^{H}\mathbf{H}_{j,j+1}\mathbf{V}_{j+1} \mathbf{s}_{j+1}}} + \tilde{\mathbf{n}}_{j} \nonumber\\
	&= \underset{\textrm{desired signal + XCI}}{\underbrace{\mathbf{U}_{j}^{H}\sum\limits_{i=1,~i\not=j+1}^{L=3}\left(\tilde{\mathbf{H}}_{j,i} + \tilde{\mathbf{E}}_{j,i}\right)\mathbf{V}_{i} \mathbf{s}_{i}}} + \underset{\textrm{ICI}}{\underbrace{\mathbf{U}_{j}^{H}\left(\tilde{\mathbf{H}}_{j,j+1} + \tilde{\mathbf{E}}_{j,j+1}\right)\mathbf{V}_{j+1} \mathbf{s}_{j+1}}} + \tilde{\mathbf{n}}_{j} \nonumber \\
	&= \underset{\textrm{desired signal}}{\underbrace{\mathbf{U}_{j}^{H}\sum\limits_{i=j-1}^{j}\tilde{\mathbf{H}}_{j,i}\mathbf{V}^{[i,j]}\mathbf{c}^{[i,j]}}} + \underset{\textrm{CSI mismatch}}{\underbrace{\mathbf{U}_{j}^{H}\sum\limits_{i=1}^{L=3}\tilde{\mathbf{E}}_{j,i}\mathbf{V}_{i}\mathbf{s}_{i}}} \nonumber\\
	&~~~+ 
	\underset{\textrm{XCI}}{\underbrace{\mathbf{U}_{j}^{H}\sum\limits_{i=j-1}^{j}\tilde{\mathbf{H}}_{j,i}\mathbf{V}^{[i,l\{l\not=j\}]}\mathbf{c}^{[i,l\{l\not=j\}]}}} + \underset{\textrm{ICI = 0}}{\underbrace{\cancel{\mathbf{U}_{j}^{H}\tilde{\mathbf{H}}_{j,j+1}\mathbf{V}_{j+1} \mathbf{s}_{j+1}}}} + \tilde{\mathbf{n}}_{j},~\forall j\in L.
\end{align}

With this in mind, we determine the data rate achievable at the user of interest as shown in \eqref{sumrate}, where  $\mathbf{J}_{j}$ in \eqref{Interference} indicates the interference terms in \eqref{Y3}.

\begin{figure}[b]
	\centering
	\includegraphics[width=0.5\columnwidth]{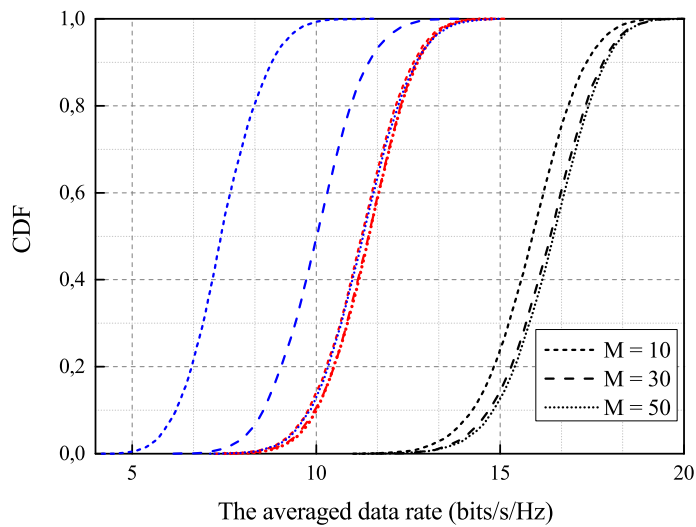}
	\caption{CDF curves with low (black), medium (red) and high (blue) correlations for different numbers of transmit antennas at 30 dB for the case of perfect CSI (when $\alpha \rightarrow \infty$).}
	\label{res1}
\end{figure}

\begin{figure*}[!t]
	\normalsize
	\begin{align}
		\label{sumrate}
		&\mathcal{I}_{\Sigma} = \sum\limits_{j=1}^{L=3}
		\mathcal{I}_{j} = \sum\limits_{j=1}^{L=3} \log_{2}\det\left(\mathbf{I} + \frac{\sum\limits_{i=j-1}^{j}  \mathbf{U}_{j}^{H}\tilde{\mathbf{H}}_{j,i}\mathbf{V}^{[i,j]} \mathbf{V}^{[i,j]H} \tilde{\mathbf{H}}_{j,i}^{H} \mathbf{U}_{j} }{\mathbf{J}_{j} + \sigma_{\tilde{n}}^{2}\mathbf{I} } \right),\\
		\label{Interference}
		&\textrm{where}~\mathbf{J}_{j} = \sum\limits_{i=j-1}^{j}  \mathbf{U}_{j}^{H}\tilde{\mathbf{H}}_{j,i}\mathbf{V}^{[i,l\{l\not=j\}]} \mathbf{V}^{[i,l\{l\not=j\}]H} \tilde{\mathbf{H}}_{j,i}^{H} \mathbf{U}_{j} 
		+ \sum\limits_{k=1}^{L=3}  \mathbf{U}_{j}^{H}\tilde{\mathbf{E}}_{j,k}\mathbf{V}_{k} \mathbf{V}_{k}^{H} \tilde{\mathbf{E}}_{j,k}^{H} \mathbf{U}_{j}.
	\end{align}
	\hrulefill
\end{figure*} 

\section{Simulation Results}
In this section, we present our results using Monte Carlo simulations with $10^5$ trials to investigate the impact of antenna correlation and CSI mismatch on the network performances. The simulations assume a Gaussian modulation and frequency flat fading which is designed according to \eqref{Hr}. The users are distributed as it was described in the system model. To make a fair comparison, the total transmit power at BS is constrained to unity irrespective of the number of transmit antennas. To create different classes of users with various numbers of antennas, we assume that BS $i$ transmits the $i$ data streams to the corresponding user $i$, and only one data stream to the collateral user of interest. Therefore, we have three receivers deployed with three, four and five antennas, respectively. 
		
According to \cite{correlation}, we examine three correlation regimes, namely, the low, medium and high correlations. The low correlation mode indicates the scenario with no correlation with sufficient spacing ($\ge \lambda/2$) between the antennas. According to \eqref{Hcorr}, the medium and high correlation regimes at the transmitter side can be modeled by $r = 0.3$ and $r = 0.9$, respectively; however, the same modes at the receiver side can be presented by $r = 0.9$. 
		
In Fig. \ref{res1}, we evaluate the cumulative distribution function (CDF) of the achievable data rate for every user in the assumed system model. The data rates are calculated using the proposed scheme under different antenna correlation modes. For the sake of clarity, we consider the average data rate achievable by the network. The observation point is 30 dB. We deploy transmitters with ten, thirty and fifty antennas to estimate the potential benefit attainable from the LS-MIMO scenario. As we can see, for the case of low correlation, the probability of attaining a higher data rate increases as the number of antennas at the BS goes up. Regarding the medium correlation case, we observe severe degradation in the achievable data rate, and the various numbers of transmit antennas do not seem to have much difference; however, a different antenna deployment still matters as it is shown in the inset figure (see Fig. \ref{res1}, where lines from top to bottom refer to the cases when transmitter is equipped with 10, 30 and 50 antennas, respectively). Finally, we consider the high correlation mode presented in blue lines with which produces a significant loss in the achievable data rate. For $M = 10$, $M = 30$ and $M = 50$ antennas deployed at the transmitter side, the average data rate of $8.8$, $11.4$ and $12.75$ bits/s/Hz are achievable with a probability of $90 \%$. It is worth to note that the deployment of more antennas allow us to overcome the high correlation, and accordingly, the system can be treated as though it is experiencing medium correlation. With this in mind, in the next simulation, we consider only the network case where all the BSs are equipped with fifty antennas. 
		
\begin{figure}[t]
	\centering
	\includegraphics[width=0.5\columnwidth]{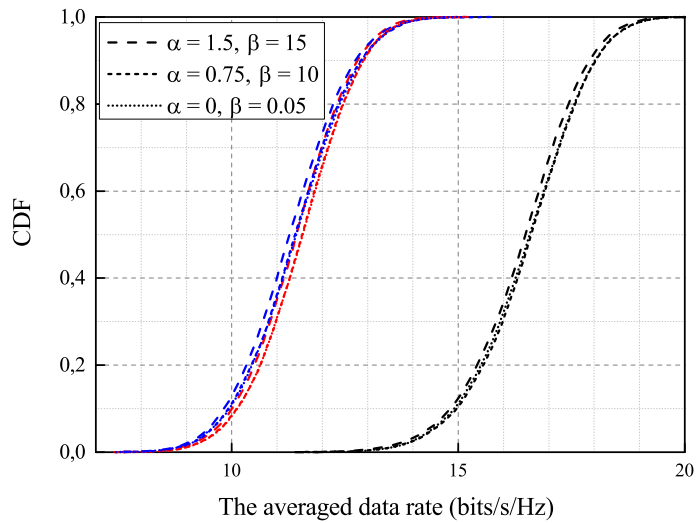} 
	\caption{CDF curves with low (black), medium (red) and high (blue) correlations for different CSI mismatch scenarios with $M=50$ antennas per BS at 30 dB.}
	\label{res2}
\end{figure}
Next, we want to investigate the combined effect of CSI mismatch and antenna correlation on the achievable data rates. The observation point is 30 dB. As shown in Fig. \ref{res2}, the case of ($\alpha = 1.5, \beta = 15$) performs worse than the other two scenarios of the CSI acquisition. The CSI mismatch cases given by ($\alpha = 0.75, \beta = 10$) and ($\alpha = 0, \beta = 0.05$) act in a similar way under the medium and high correlation regimes; however, for low correlation, the network with ($\alpha = 0, \beta = 0.05$) slightly outperforms the one modeled by ($\alpha = 0.75, \beta = 10$). This result leads to the realization that the SNR-dependent and SNR-independent CSI acquisition scenarios, ($\alpha = 0, \beta = 0.05$) and  ($\alpha = 0.75, \beta = 10$), do not differ from each other in the cases of medium and high antenna correlations.

It is worth mentioning that the provided results in Fig. 4 and 5 also provide a representative insight into how many DoFs can be achieved under the considered network scenario. Eq. \eqref{eta} can be modified to calculate the average number of DoFs as follows
\begin{equation}
\textrm{DoF} = \frac{\mathcal{I}(SNR)}{\log_{2}(SNR)},
\end{equation}
when the averaged sum rate values are scaled by the SNR value (the observation point is chosen to be 30 dB).	
		
\begin{figure}[h]
	\centering
	\includegraphics[width=0.5\columnwidth]{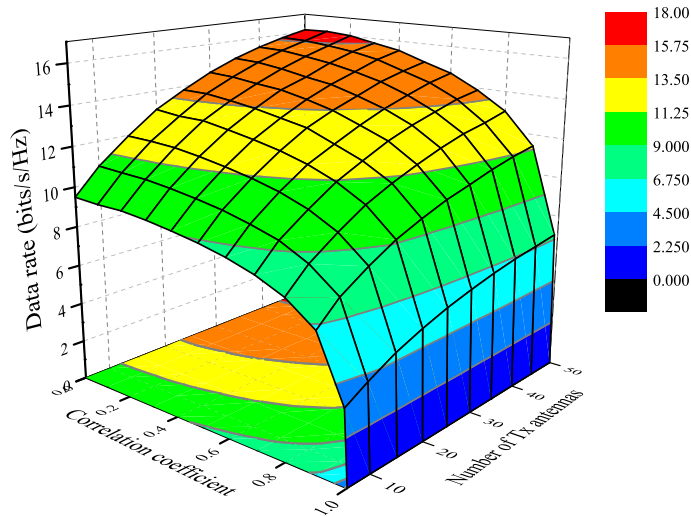} 
	\caption{The achievable data rate as a function of the correlation coefficient and number of transmit antennas.}
	\label{res8}
\end{figure}
Finally, we provide a 3D plot in Fig. \ref{res8} presenting the data rate achievable at 30 dB as a function of the number of transmit antennas and correlation coefficient. The correlation at the transmitter side is assumed to be high and modeled by using the exponential model as in \eqref{corrExp} ($r = 0.9$). Accordingly, the antenna correlation at the receiver side is given as in \eqref{corrUni} and simulated within the range $[0, 1]$. As we can see, the achievable data rate increases as the number of transmit antennas goes up, while the data rate decreases as the correlation coefficient increases. 
		
\section{Conclusion}
In this work, we considered the compounded MIMO BC network scenario and proposed a generalized transmit BF design that accounts for different user classes. We analysed the feasibility conditions of IA and presented the DoF region analysis which were then utilized in design of an algorithm to define the minimum antenna configuration to achieve a required number of data streams in the network. Moreover, we investigated the impact of spatial antenna correlation under various CSI acquisition scenarios. Finally, the proposed scheme was examined in traditional and LS-MIMO systems under different channel scenarios. It was shown that the performance obtained for the latter case indicates that deploying more antennas makes it possible to overcome the impact of high correlation by careful manipulation of the antenna array.
	
		
		\ifCLASSOPTIONcaptionsoff
		\newpage
		\fi

\end{document}